# Diffusions and anomalous diffusions of charged particles in the plateau regime of toroidal plasma

Hong Wang[1], Jiulin Du[(a)]

[1]*Department of Physics, School of Science, Tianjin University, Tianjin 300072, China*
[2]*Department of Physics, School of Science, Tianjin University, Tianjin 300072, China*

**Abstract**   The diffusions and anomalous diffusions of charged particles in the plateau regime of toroidal plasma with a Maxwellian velocity distribution are studied. The transport theory of plasma is discussed under the toroidal coordinate system and the magnetic field is assumed to be axisymmetric. Based on the diffusion flux obtained in the toroidal plasma, we derive the diffusion coefficient tensor and thermal diffusion coefficient tensor, where two angular diffusion coefficients are the anomalous diffusions, depending on the radial position and having nothing to do with the magnetic field. All the radial transport coefficients and the normal angular coefficients are inversely proportional to square of the gyrofrequency and so to square of the magnetic induction. The numerical analyses show clearly dependences of these complex transport coefficients on the temperature, the magnetic fluid stability safe factor and the effective collision frequency, respectively.

## 1 Introduction

The toroidal plasma is the tokamak plasma which uses a strong toroidal magnetic field to contain high temperature plasma within the torus. The transport properties of charged particles in the toroidal plasma are very complex. In macroscopic inhomogeneous medium, the random collisions of particles can be balanced by the free flow of individual particles. Since the medium is also affected by possible external force fields, such as gravity and electromagnetic force etc., the plasma system may eventually reach a quasi-steady state through the mutual restriction of collisions, flows and responses to the external forces. The quasi-equilibrium plasma system is an axisymmetric toroidal equilibrium, filled with electrons and ions [1-3]. If the charged particle has its position **r** and velocity **v**, and the equilibrium probability distribution function $f$ (**r**, **v**) is a Maxwellian velocity distribution, then we have that

$$f_\alpha(\mathbf{r},\mathbf{v}) = n_\alpha(\mathbf{r}) \left( \frac{m_\alpha}{2\pi k_B T_\alpha(\mathbf{r})} \right)^{\frac{3}{2}} \exp\left( -\frac{m_\alpha \mathbf{v}^2}{2k_B T_\alpha(\mathbf{r})} \right), \qquad (1)$$

where the subscript $\alpha = e, i$ for electrons and ions, $n_\alpha$, $T_\alpha$ and $m_\alpha$ are respectively the number density, temperature and mass of the $\alpha$th composition of the plasma, and $k_B$ is Boltzmann constant.

It is well known that the presence of magnetic field strongly affects the behavior of ionized gas and plasma. The movements of particles in most space and laboratory plasma are restricted by the magnetic fields. This conclusion has been confirmed in magnetic fusion experiments [4, 5] and planetary magnetosphere [6]. The researches on the transport processes including particles, momentum and energy in the plasma is one of the main research contents of the collision plasma physics [2]. The particle and energy transports in the plasma system that is confined by a toroidal magnetic field include the classical transport, the neoclassical transport and the anomalous (or turbulent) transport. The classical and the neoclassical transport are both caused by the Coulomb collisions of charged particles, while the anomalous transport is often the turbulent fluctuations driven by a various of instability in the confined plasma [6-11]. The particle and energy transport fluxes observed in the most fusion experiments that exceed those predicted with the neoclassical theory are called anomalous transports [9]. In the collisional transport processes, the neoclassical transport concentrates more on the drifting motion of the guiding center in a toroidal magnetic configuration [10], while the classical transport mainly studies the cyclotron motion of particles

[11].

In the past about 50 years, the classic plasma collision theories have been developed on the basis of the Chapman-Enskog-Grad formalism [12,13]. Since the transport rate of controlled nuclear fusion is significant, theoretical and experimental investigation of the magnetically confined plasma transport processes have been widely promoted [14]. However, the error between theoretical results and experimental observations differs sometimes by several orders of magnitude. The difficulty may come from two aspects. On the one hand, the collision in fully ionized plasma can no longer be simply regarded as a two-body collision since the coulomb force is a long-range interaction [15-17]. On the other hand, the response of charged particles to the magnetic field may cause the particle drift effects and particle trapping, which hinders the prediction of transport phenomena in the plasma to a certain extent.

When we explore the particle and energy balance, it is essential to consider the collisional diffusion of impurities in the tokamak plasma and the subsequent ion diffusion and energy transport [18]. In previous studies, the fluid equation of the neoclassical transport [7,19] provides a powerful framework for estimating the effects of impurity transport [20], bipolar electric field [21], plasma rotation [22], current and pulse source [7,23]. In astrophysics and laboratory plasmas, the relative magnitude of the collision frequency $\nu$ and the transition frequency $\Omega_T = v_T/\gamma R_0$ dominates the physical mechanism of neoclassical transport [2], where $v_T = \sqrt{2k_B T/m}$ is the thermal velocity of the particles with temperature $T$ and mass $m$, $\gamma$ is the magnetic fluid stability safety factor that is related to the magnetic field configuration, $R_0$ is the curvature radius of the magnetic field lines, which characterizes the distance between the ring of the toroidal magnetic field and the central axis. If the plasma collision frequency $\nu$ is much higher than the transition frequency $\Omega_T$, namely,

$$\frac{S}{\lambda} \sim \frac{\nu}{v_T/\gamma R_0} \gg 1 , \qquad (2)$$

the particles is in the Pfirsch-Schluter or fluid regime [24]. The further physical meaning is that the mean free path $\lambda = v_T/\nu$ is much smaller than the parallel distance $S \sim qR_0$ around the flux surface. In this case, the Braginskii fluid equation is reasonably used to analyze its transport process [25]. On the contrary, if $\nu/(v_T/\gamma R_0) \ll 1$, it referred to as the banana-plateau regime. In this case, the particle orbits are completed and the short mean free path is no longer valid. In other words, a kinetic description is indispensable in the long mean-free path regimes, plateau and banana [26]. In the laboratory plasma, the core of a tokamak is usually in the banana-plateau regime. And if the aspect ratio (namely, the ratio of the edge radius $a$ constrained by the toroidal magnetic field configuration to the distance $R_0$ between the ring of the toroidal magnetic field and the central axis) is large, i.e., $\varepsilon \ll 1$, the banana-plateau regime is subdivided into two regimes: i.e., the plateau regime [27] when

$$\varepsilon^{3/2} \ll \frac{\nu}{v_T/\gamma R_0} \ll 1 , \qquad (3)$$

and the banana regime [28] when

$$\frac{\nu}{v_T/\gamma R_0} \ll \varepsilon^{3/2} . \qquad (4)$$

In the early years, the transport phenomenon in the banana regime attracted the interest of many researchers due to the particularity of the banana orbit. For example, Taguchi used a new approximate collision operator to calculate the thermal conductivity and distribution function of ions in the banana regime for an axisymmetric toroidal plasma of arbitrary aspect ratio [29]. Wang *et al.* used the strict Chapman-Enskog-like method for the first time to determine the polar changes of the parallel viscous force and the thermal viscous force [30]. Hirshman *et al.* established a comprehensive neoclassical transport theory for multi-species plasma in the banana regime with a low collision frequency [7]. In these references, we find that the collision frequency plays an important role in the transport of charged particles. Due to the inhomogeneity of magnetic field, the particle trajectory in the plateau regime is a complex, spiral and unclosed curve. Therefore, the research on the transport of charged particles in the plateau regime is still imperfect. In this work, we study the diffusions and anomalous diffusions of charged particles in the plateau regime of the toroidal plasma, and analyze the effects of the magnetic field, the temperature, the safe factor and the effective collision frequency $v_{eff} = v/\varepsilon$ on the transport coefficients [2].

The paper is organized as follows. In Sec. 2, we use the kinetic transport equation of the toroidal plasma system to study the velocity distribution function of charged particles confined by the toroidal axisymmetric magnetic field in the quasi-steady state. In Sec. 3, we study the total diffusion flux vector in the plasma system and then find the transport coefficient tensors. In Sec. 4, we study the derived transport coefficients numerically. Finally, in Sec. 5 we give the conclusion.

## 2 The transport theory of charged particles in the toroidal plasma

In the self-consistent electromagnetic field that is generated by the external electromagnetic field and the internal current and charge in the plasma, the governing transport equation for the velocity distribution function can be expressed [1,2, 31-34] as

$$\frac{\partial f_\alpha}{\partial t} + \mathbf{v} \cdot \frac{\partial f_\alpha}{\partial \mathbf{r}} + \frac{Q_\alpha}{m_\alpha}\left(\mathbf{E} + \frac{1}{c}\mathbf{v} \times \mathbf{B}\right) \cdot \frac{\partial f_\alpha}{\partial \mathbf{v}} = C(f_\alpha) , \qquad (5)$$

where $f_\alpha = f_\alpha(\mathbf{r}, \mathbf{v}, t)$ is a single-particle velocity distribution function at time $t$, velocity $\mathbf{v}$, and position $\mathbf{r}$, $Q_\alpha$ is the charge of the $\alpha$-th species, $\mathbf{B}$ is the magnetic induction intensity, and $\mathbf{E}$ is the electric field intensity, $C(f_\alpha)$ is the collision term in the kinetic theory [2]. When the plasma is constrained by a toroidal strong magnetic field, the movement of the charged particles contains the rotating motion around the guiding center, the movement in the direction of magnetic field line, and the drift motion perpendicular to the magnetic field. For simplicity, the distribution function of charged particle can be regarded as the distribution function of the guiding center. The motion of the guiding center particle is the superposition of the longitudinal motion along the twisted magnetic field line and the drift motion caused by the non-uniformity of the magnetic field, and then the position of the guiding center is $\mathbf{r}_g$ and the velocity of the guiding center is $\mathbf{v}_g = \mathbf{v}_{//} + \mathbf{v}_D$, where we assume that the drift velocity $\mathbf{v}_D$ is a first-order small amount compared with velocity $\mathbf{v}_{//}$ that is parallel to the magnetic field [38]. Thus the evolution equation of the guiding center velocity over time is expressed [2,34] as

$$\frac{\partial f_\alpha}{\partial t} + \left(\mathbf{v}_{//} + \mathbf{v}_D\right) \cdot \frac{\partial f_\alpha}{\partial \mathbf{r}_g} + \dot{\mathbf{v}}_g \cdot \frac{\partial f_\alpha}{\partial \mathbf{v}_g} = \left(\frac{\partial f_\alpha}{\partial t}\right)_{coll} . \qquad (6)$$

Eq.(6) is also called the drift kinetic equation when the particles move in a toroidal magnetic field. The kinetic energy $W_\alpha$ and the magnetic moment $\mu_\alpha$ of the $\alpha$-th species are defined respectively as

$$W_\alpha = \frac{m_\alpha}{2}\left(v_{//}^2 + v_\perp^2\right), \quad \mu_\alpha = m_\alpha v_\perp^2 / 2B , \qquad (7)$$

where $\mathbf{v}_\perp$ is the velocity perpendicular to the magnetic field. Therefore, correspondingly, the Maxwellian velocity distribution function of the $\alpha$-th species can also be expressed by the kinetic energy $W_\alpha$ and magnetic moment $\mu_\alpha$ as

$$f_\alpha = f_\alpha\left(\mathbf{r}_g, W_\alpha, \mu_\alpha, t\right) . \qquad (8)$$

According to the invariance of Liouville equation in the transformation of motion variables

[35], the drift kinetic Eq. (6) can be rewritten for the distribution (8) as

$$\frac{\partial f_\alpha}{\partial t} + \frac{\partial f_\alpha}{\partial W_\alpha}\frac{dW_\alpha}{dt} + \frac{\partial f_\alpha}{\partial \mu_\alpha}\frac{\partial \mu_\alpha}{\partial t} + \mathbf{v}_g \cdot \frac{\partial f_\alpha}{\partial \mathbf{r}_g} = \left(\frac{\partial f_\alpha}{\partial t}\right)_{coll}. \quad (9)$$

If we study a local problem, meaning [2,15,32-34] that

$$|h \cdot \nabla B| \ll B, \; |\mathbf{r}_c \cdot \nabla B| \ll B, \; \left|\varpi_c^{-1} \cdot \frac{\partial B}{\partial t}\right| \ll 1, \quad (10)$$

the magnetic field can be approximately considered as invariant, where $r_c$, $h$ and $\varpi_c$ are respectively the radius of gyration, the pitch and the gyration frequency of the $\alpha$-th species moving in a toroidal magnetic field. In this physical case, both the magnetic moment $\mu_\alpha$ and the total kinetic $W_\alpha$ of the particles can be approximately constant. In the local quasi-steady state, $\partial f_\alpha / \partial t = 0$, the drift kinetic Eq. (9) can be simplified as

$$\mathbf{v}_g \cdot \frac{\partial f_\alpha}{\partial \mathbf{r}_g} = \left(\frac{\partial f_\alpha}{\partial t}\right)_{coll}. \quad (11)$$

As usual, in the first-order approximation of Chapman-Enskog expansion [34,36,37], we write the velocity distribution function in the following form,

$$f_\alpha = f_{M,\alpha}^{(0)} + f_\alpha^{(1)}, \quad (12)$$

where $f_\alpha^{(1)}$ is the first-order small disturbance about the equilibrium distribution function, the Maxwellian distribution $f_{M,\alpha}^{(0)}$. After the independent variables $(\mathbf{r}_g, \mathbf{v}_g)$ are replaced by $(W_\alpha, \mu_\alpha)$, the equilibrium distribution function $f_{M,\alpha}^{(0)}$ is changed into the following form,

$$f_{M,\alpha}^{(0)}(\mathbf{r}_g, W_\alpha, \mu_\alpha) = n_\alpha(\mathbf{r}_g)\left(\frac{m_\alpha}{2\pi k_B T_\alpha(\mathbf{r}_g)}\right)^{\frac{3}{2}} \exp\left(-\frac{(W_\alpha - \mu_\alpha B(\mathbf{r}_g))}{k_B T_\alpha(\mathbf{r}_g)}\right). \quad (13)$$

where $\mathbf{v}_g \approx \mathbf{v}_{//}$ has been used because of $\mathbf{v}_D \ll \mathbf{v}_{//}$.

We assume that Larmor radius $\rho$ is small as compared with the macroscopic scale length $L$, namely, $\delta \equiv \rho / L \ll 1$. If the velocity of the electrical drift is ignored, it is considered that $\mathbf{v}_D$ is only the sum of the gradient drift and the curvature drift velocity caused by the toroidal inhomogeneity of the magnetic field,

$$\mathbf{v}_D = \frac{1}{\varpi_c R_0}\left(\mathbf{v}_{//}^2 + \frac{1}{2}\mathbf{v}_\perp^2\right), \quad (14)$$

where $\varpi_c = |Q_\alpha|B/m_\alpha$ is the gyrofrequency of particles in the toroidal magnetic field. In the toroidal coordinate reference system $(\mathbf{e}_\rho, \mathbf{e}_\theta, \mathbf{e}_\xi)$, the parallel velocity $\mathbf{v}_{//}$ and the drift velocity $\mathbf{v}_D$ of the particle are, respectively [2,34],

$$\mathbf{v}_{//} = \left(0, \frac{v_{//}\rho_g}{\gamma R_0}, v_{//}\right), \; \mathbf{v}_D = (v_D \sin\theta, v_D \cos\theta, 0), \quad (15)$$

where $\gamma = \frac{B_\varphi}{B_\theta}\frac{a}{R_0}$ is the magnetic fluid safety factor which is related to the magnetic restraint device. Therefore, the velocity of the guiding center particle is that

$$\mathbf{v}_g = \mathbf{v}_{//} + \mathbf{v}_D = \begin{pmatrix} 0 \\ \frac{v_{//}\rho_g}{\gamma R_0} \\ v_{//} \end{pmatrix} + \begin{pmatrix} v_D \sin\theta \\ v_D \cos\theta \\ 0 \end{pmatrix} = \begin{pmatrix} v_D \sin\theta \\ \frac{v_{//}\rho_g}{\gamma R_0} + v_D \cos\theta \\ v_{//} \end{pmatrix}. \quad (16)$$

In the toroidal coordinate reference system, the gradient formula of a scalar $A$ is that [2,8]

$$\nabla A = \frac{\partial A}{\partial \rho} \mathbf{e}_\rho + \frac{1}{\rho} \frac{\partial A}{\partial \theta} \mathbf{e}_\theta + \frac{1}{R_0 + \rho \cos\theta} \frac{\partial A}{\partial \xi} \mathbf{e}_\xi . \tag{17}$$

If we consider the case of an axisymmetric magnetic field, we have that $\partial/\partial \xi = 0$. And because $f_{M,\alpha}^{(0)}$ has no relationship with the angular coordinates, we get that

$$\frac{\partial f_{M,\alpha}}{\partial \boldsymbol{\rho}_g} = \left[ \frac{\partial f_{M,\alpha}^{(0)}}{\partial \rho_g} + \frac{\partial f_{M,\alpha}^{(1)}}{\partial \rho_g}, \frac{1}{\rho_g}\left( \frac{\partial f_{M,\alpha}^{(0)}}{\partial \theta} + \frac{\partial f_{M,\alpha}^{(1)}}{\partial \theta} \right), 0 \right], \tag{18}$$

and so,

$$\mathbf{v}_g \cdot \frac{\partial f_{M,\alpha}}{\partial \boldsymbol{\rho}_g} = v_D \sin\theta \left( \frac{\partial f_{M,\alpha}^{(0)}}{\partial \rho_g} + \frac{\partial f_{M,\alpha}^{(1)}}{\partial \rho_g} \right) + \left( \frac{v_{//}}{\gamma R_0} + v_D \cos\theta \frac{1}{\rho_g} \right)\left( \frac{\partial f_{M,\alpha}^{(0)}}{\partial \theta} + \frac{\partial f_{M,\alpha}^{(1)}}{\partial \theta} \right). \tag{19}$$

In Eq. (19), only the first-order of the terms is retained so that

$$\mathbf{v}_g \cdot \frac{\partial f_{M,\alpha}}{\partial \boldsymbol{\rho}_g} = v_D \sin\theta \frac{\partial f_{M,\alpha}^{(0)}}{\partial \rho_g} + \left( \frac{v_{//}}{\gamma R_0} + v_D \cos\theta \frac{1}{\rho_g} \right) \frac{\partial f_{M,\alpha}^{(0)}}{\partial \theta} + \frac{v_{//}}{\gamma R_0} \frac{\partial f_{M,\alpha}^{(1)}}{\partial \theta} . \tag{20}$$

In Eq.(13), the collision term uses the Krook collision model [2,38], i.e.,

$$\left( \frac{\partial f_{M,\alpha}}{\partial t} \right)_{coll} = -\frac{1}{\tau_\alpha}\left( f_{M,\alpha} - f_{M,\alpha}^{(0)} \right) = -\nu_{\alpha,eff} f_{M,\alpha}^{(1)} , \tag{21}$$

where $\tau_\alpha$ is the average collision time and $\nu_{\alpha,\,eff}$ is the effective collision frequency of the *α*-th species. Then the drift kinetic Eq. (11) becomes

$$v_D \sin\theta \frac{\partial f_{M,\alpha}^{(0)}}{\partial \rho_g} + \left( \frac{v_{//}}{\gamma R_0} + v_D \cos\theta \frac{1}{\rho_g} \right) \frac{\partial f_{M,\alpha}^{(0)}}{\partial \theta} + \frac{v_{//}}{\gamma R_0} \frac{\partial f_{M,\alpha}^{(1)}}{\partial \theta} = -\nu_{\alpha,eff} f_{M,\alpha}^{(1)} . \tag{22}$$

By solving Eq. (22), we can obtain the first-order approximation of Chapman-Enskog expansion for the velocity distribution function,

$$f_{M,\alpha}^{(1)} = -\frac{v_{//}}{\gamma R_0} \frac{1}{\nu_{\alpha,eff}} \frac{\partial f_{M,\alpha}^{(0)}}{\partial \theta} - \left( \frac{v_{//}^2}{\gamma^2 R_0^2} + \nu_{\alpha,eff}^2 \right)^{-1} \left( -\frac{v_{//}}{\gamma R_0} \frac{\partial f_{M,\alpha}^{(0)}}{\partial \rho_g} + \frac{\nu_{\alpha,eff}}{\rho_g} \frac{\partial f_{M,\alpha}^{(0)}}{\partial \theta} \right) v_D \cos\theta$$

$$-\left( \frac{v_{//}^2}{\gamma^2 R_0^2} + \nu_{\alpha,eff}^2 \right)^{-1} \left( \frac{v_{//}}{\gamma R_0} \frac{1}{\rho_g} \frac{\partial f_{M,\alpha}^{(0)}}{\partial \theta} + \nu_{\alpha,eff} \frac{\partial f_{M,\alpha}^{(0)}}{\partial \rho_g} \right) v_D \sin\theta . \tag{23}$$

According to that

$$\frac{\partial f_{M,\alpha}^{(0)}}{\partial \boldsymbol{\rho}_g} = \frac{\partial f_{M,\alpha}^{(0)}}{\partial \rho_g} \mathbf{e}_\rho + \frac{1}{\rho_g} \frac{\partial f_{M,\alpha}^{(0)}}{\partial \theta} \mathbf{e}_\theta , \tag{24}$$

and the Maxwellian distribution (1), we have that

$$\frac{\partial f_{M,\alpha}^{(0)}}{\partial \rho_g} = \frac{f_{M,\alpha}^{(0)}}{n_\alpha} \frac{\partial n_\alpha}{\partial \rho_g} + \left( \frac{m_\alpha \mathbf{v}_g^2}{k_B T_\alpha} - 3 \right) \frac{f_{M,\alpha}^{(0)}}{2T_\alpha} \frac{\partial T_\alpha}{\partial \rho_g} ,$$

$$\frac{\partial f_{M,\alpha}^{(0)}}{\partial \theta} = \frac{f_{M,\alpha}^{(0)}}{n_\alpha} \frac{\partial n_\alpha}{\partial \theta} + \left( \frac{m_\alpha \mathbf{v}_g^2}{k_B T_\alpha} - 3 \right) \frac{f_{M,\alpha}^{(0)}}{2T_\alpha} \frac{\partial T_\alpha}{\partial \theta} . \tag{25}$$

## 3 The diffusion flux and the diffusion coefficients in the toroidal plasma

The magnetic field configuration of Tokamak is axisymmetric [8], $\mathbf{B} = B_\theta \mathbf{e}_\theta + B_\xi \mathbf{e}_\xi$, so for a large aspect ratio $|B_\xi| \gg |B_\theta|$, the magnetic field strength $B \approx |B_\xi|$. Since the plasma is anisotropic, the diffusion coefficient and thermal diffusion coefficient are both second-order tensors. According to Onsager relation, the diffusion flux vector of the α-th species is related to the particle number density gradient and the temperature gradient by the equation [15,31,33,40],

$$\mathbf{J}_{D,\alpha} = -\mathbf{D}_\alpha \nabla n_\alpha - n_\alpha \mathbf{D}_{T\alpha} \nabla T_\alpha ,\qquad (26)$$

where $\mathbf{D}_\alpha$ is the diffusion coefficient tensor, $\mathbf{D}_{T\alpha}$ is the thermal diffusion coefficient tensor. In the toroidal coordinate reference system, they are expressed as matrixes,

$$\mathbf{D}_\alpha = \begin{pmatrix} D^\rho_{\alpha,\rho} & D^\rho_{\alpha,\theta} & D^\rho_{\alpha,\xi} \\ D^\theta_{\alpha,\rho} & D^\theta_{\alpha,\theta} & D^\theta_{\alpha,\xi} \\ D^\xi_{\alpha,\rho} & D^\xi_{\alpha,\theta} & D^\xi_{\alpha,\xi} \end{pmatrix}, \text{ and } \mathbf{D}_{T\alpha} = \begin{pmatrix} D^\rho_{T\alpha,\rho} & D^\rho_{T\alpha,\theta} & D^\rho_{T\alpha,\xi} \\ D^\theta_{T\alpha,\rho} & D^\theta_{T\alpha,\theta} & D^\theta_{T\alpha,\xi} \\ D^\xi_{T\alpha,\rho} & D^\xi_{T\alpha,\theta} & D^\xi_{T\alpha,\xi} \end{pmatrix}, \qquad (27)$$

where $D^j_{\alpha,i}\left(D^j_{T\alpha,i}\right)$ represents the diffusion (thermal diffusion) coefficient in the $j$ direction caused by the non-uniformity in the $i$ direction ($i,j = \rho_g, \theta, \xi$). Because the toroidal plasma under consideration is axisymmetric, we conclude that $D^\rho_{\alpha,\xi}=D^\theta_{\alpha,\xi}=D^\xi_{\alpha,\xi}=0$ and $D^\rho_{T\alpha,\xi}=D^\theta_{T\alpha,\xi}=D^\xi_{T\alpha,\xi}=0$. Thus, the diffusion flux vector (26) can be written as

$$\mathbf{J}_{D,\alpha} = -\begin{pmatrix} D^\rho_{\alpha,\rho} & D^\rho_{\alpha,\theta} & 0 \\ D^\theta_{\alpha,\rho} & D^\theta_{\alpha,\theta} & 0 \\ D^\xi_{\alpha,\rho} & D^\xi_{\alpha,\theta} & 0 \end{pmatrix} \begin{pmatrix} \dfrac{\partial n_\alpha}{\partial \rho_g} \\ \dfrac{1}{\rho_g}\dfrac{\partial n_\alpha}{\partial \theta} \\ 0 \end{pmatrix} - n_\alpha \begin{pmatrix} D^\rho_{T\alpha,\rho} & D^\rho_{T\alpha,\theta} & 0 \\ D^\theta_{T\alpha,\rho} & D^\theta_{T\alpha,\theta} & 0 \\ D^\xi_{T\alpha,\rho} & D^\xi_{T\alpha,\theta} & 0 \end{pmatrix} \begin{pmatrix} \dfrac{\partial T_\alpha}{\partial \rho_g} \\ \dfrac{1}{\rho_g}\dfrac{\partial T_\alpha}{\partial \theta} \\ 0 \end{pmatrix}$$

$$= -\begin{pmatrix} D^\rho_{\alpha,\rho}\dfrac{\partial n_\alpha}{\partial \rho_g} + D^\rho_{\alpha,\theta}\dfrac{1}{\rho_g}\dfrac{\partial n_\alpha}{\partial \theta} + n_\alpha D^\rho_{T\alpha,\rho}\dfrac{\partial T_\alpha}{\partial \rho_g} + n_\alpha D^\rho_{T\alpha,\theta}\dfrac{1}{\rho_g}\dfrac{\partial T_\alpha}{\partial \theta}, \\ D^\theta_{\alpha,\rho}\dfrac{\partial n_\alpha}{\partial \rho_g} + D^\theta_{\alpha,\theta}\dfrac{1}{\rho_g}\dfrac{\partial n_\alpha}{\partial \theta} + n_\alpha D^\theta_{T\alpha,\rho}\dfrac{\partial T_\alpha}{\partial \rho_g} + n_\alpha D^\theta_{T\alpha,\theta}\dfrac{1}{\rho_g}\dfrac{\partial T_\alpha}{\partial \theta}, \\ D^\xi_{\alpha,\rho}\dfrac{\partial n_\alpha}{\partial \rho_g} + D^\xi_{\alpha,\theta}\dfrac{1}{\rho_g}\dfrac{\partial n_\alpha}{\partial \theta} + n_\alpha D^\xi_{T\alpha,\rho}\dfrac{\partial T_\alpha}{\partial \rho_g} + n_\alpha D^\xi_{T\alpha,\theta}\dfrac{1}{\rho_g}\dfrac{\partial T_\alpha}{\partial \theta} \end{pmatrix} \qquad (28)$$

On the other hand, the density flux vector of the α-th species is defined by the velocity distribution function [2,31,39] as

$$\mathbf{J}_{D,\alpha} = n_\alpha <\mathbf{v}_g> = \int \mathbf{v}_g f_{M,\alpha}\left(\boldsymbol{\rho}_g, \mathbf{v}_g\right) d\mathbf{v}_g = \int \mathbf{v}_g \left(f^{(0)}_{M,\alpha} + f^{(1)}_{M,\alpha}\right) d\mathbf{v}_g . \qquad (29)$$

Because $f^{(0)}_{M,\alpha}$ is an even function of $\mathbf{v}_g$, the first term, the integral for $f^{(0)}_{M,\alpha}$, is zero, and then the diffusion flux vector becomes

$$\mathbf{J}_{D,\alpha} = \int \mathbf{v}_g f^{(1)}_{M,\alpha} d\mathbf{v}_g = \int \left(v_D \sin\theta \mathbf{e}_\rho + \left(\frac{v_\parallel \rho_g}{\gamma R_0} + v_D \cos\theta\right)\mathbf{e}_\theta + v_\parallel \mathbf{e}_\xi\right) f^{(1)}_{M,\alpha} d\mathbf{v}_g ,$$

$$= J^\rho_{D,\alpha}\mathbf{e}_\rho + J^\theta_{D,\alpha}\mathbf{e}_\theta + J^\xi_{D,\alpha}\mathbf{e}_\xi \qquad (30)$$

where, the diffusion flux component in the $\mathbf{e}_\rho$-direction is given by

$$J^\rho_{D,\alpha} = \iiint_{\substack{0<v_\parallel<\infty,\\-\pi<\theta<\pi,\\0<\xi<2\pi}} v_D \sin\theta f^{(1)}_{M,\alpha} dv_\parallel d\theta d\xi = -\iiint_{\substack{0<v_\parallel<\infty,\\-\pi<\theta<\pi,\\0<\xi<2\pi}} \frac{v_\parallel}{\gamma R_0} \frac{v_D \sin\theta}{v_{\alpha,\mathit{eff}}} \frac{\partial f^{(0)}_{M,\alpha}}{\partial \theta} dv_\parallel d\theta d\xi$$

$$- \iiint_{\substack{0<v_\parallel<\infty,\\-\pi<\theta<\pi,\\0<\xi<2\pi}} v^2_D \sin\theta \cos\theta \frac{-\dfrac{v_\parallel}{\gamma R_0}\dfrac{\partial f^{(0)}_{M,\alpha}}{\partial \rho_g} + \dfrac{v_{\alpha,\mathit{eff}}}{\rho_g}\dfrac{\partial f^{(0)}_{M,\alpha}}{\partial \theta}}{v^2_{\alpha,\mathit{eff}} + v^2_\parallel/\gamma^2 R^2_0} dv_\parallel d\theta d\xi$$

$$-\iiint\limits_{\substack{0<v_{//}<\infty,\\-\pi<\theta<\pi,\\0\le\xi<2\pi}} v_D^2 \sin^2\theta \frac{\frac{v_{//}}{\gamma R_0}\frac{1}{\rho_g}\frac{\partial f_{M,\alpha}^{(0)}}{\partial \theta}+\nu_{\alpha,eff}\frac{\partial f_{M,\alpha}^{(0)}}{\partial \rho_g}}{\nu_{\alpha,eff}^2+v_{//}^2/\gamma^2 R_0^2} dv_{//}d\theta d\xi \; , \quad (31)$$

where, because the drift velocity is a first-order small quantity $v_D \ll v_{//}$, we consider approximately $v_g \approx v_{//}$ and $v_g^2 = v_{//}^2 + v_\perp^2 \approx v_{//}^2$, and the Maxwellian velocity distribution of the guiding center particle is reduced to be

$$f_{M,\alpha}^{(0)}(\mathbf{r}_g, \mathbf{v}_g) = n_\alpha(\mathbf{r}_g)\left(\frac{m_\alpha}{2\pi k_B T_\alpha(\mathbf{r}_g)}\right)^{\frac{3}{2}} \exp\left(-\frac{m_\alpha v_{//}^2}{2k_B T_\alpha(\mathbf{r}_g)}\right). \quad (32)$$

Since the collision frequency of the charge particles in the plateau regime is particularly small, closing to collisionless, $\nu_{\alpha,eff} \ll v_{T,\alpha}/\gamma R_0$, so the term, $\nu_{\alpha,eff}/(\nu_{\alpha,eff}^2 + v_{//}^2/\gamma^2 R_0^2)$, takes a maximum value near $v_{//} \simeq 0$. Using this limit condition and the limit formula [34],

$$\lim_{\varepsilon \to 0} \frac{\varepsilon}{\varepsilon^2 + x^2} = \pi\delta(x) \; , \quad (33)$$

the term, $\nu_{\alpha,eff}/(\nu_{\alpha,eff}^2 + v_{//}^2/\gamma^2 R_0^2)$, can be replaced by the Dirac function, $\pi\delta(v_{//}/\gamma R_0)$. In Eq.(15), because $v_D$ is very small and it is almost a constant, a reasonable approximation is usually considered [34] as

$$v_D \approx \frac{1}{\varpi_c R_0}\left(v_{//}^2 + v_\perp^2\right) \approx \frac{v_T^2}{\varpi_c R_0} = \frac{k_B T_\alpha}{m_\alpha \varpi_c R_0} \; . \quad (34)$$

After doing the above approximation and simplification, Eq.(31) can be calculated (see Appendix) as

$$J_{D,\alpha}^\rho = -2\pi^{3/2}\sqrt{\frac{2k_B T_\alpha}{m_\alpha}}\frac{\gamma}{\varpi_c^2 R_0}\frac{\partial n_\alpha}{\partial \rho_g} + 3\pi\sqrt{\frac{2\pi k_B T_\alpha}{m_\alpha}}\frac{\gamma}{\varpi_c^2 R_0}\frac{n_\alpha}{T_\alpha}\frac{\partial T_\alpha}{\partial \rho_g}$$
$$-\sqrt{\frac{2\pi k_B T_\alpha}{m_\alpha}}\frac{\gamma}{\varpi_c^2 R_0}\Gamma\left(0, \frac{m_\alpha \nu_{\alpha,eff}^2 \gamma^2 R_0^2}{2k_B T_\alpha}\right)\exp\left(\frac{m_\alpha \nu_{\alpha,eff}^2 \gamma^2 R_0^2}{2k_B T_\alpha}\right)\frac{1}{\rho_g}\frac{\partial n_\alpha}{\partial \theta}$$
$$-\sqrt{\frac{2\pi k_B}{m_\alpha T_\alpha}}\frac{\gamma}{2\varpi_c^2 R_0}\left[2-\left(3+\frac{\gamma^2 R_0^2}{k_B T_\alpha}m_\alpha \nu_{\alpha,eff}^2\right)\Gamma\left(0, \frac{m_\alpha \nu_{\alpha,eff}^2 \gamma^2 R_0^2}{2k_B T_\alpha}\right)\exp\left(\frac{m_\alpha \nu_{\alpha,eff}^2 \gamma^2 R_0^2}{2k_B T_\alpha}\right)\right]\frac{n_\alpha}{\rho_g}\frac{\partial T_\alpha}{\partial \theta} . \quad (35)$$

As compared with Eq.(28) we find the four radial coefficients, where two radial diffusion coefficients are respectively,

$$D_{\alpha,\rho}^\rho = \frac{2\pi\gamma}{\varpi_c^2 R_0}\sqrt{\frac{2\pi k_B T_\alpha}{m_\alpha}} \; , \quad (36)$$

$$D_{\alpha,\theta}^\rho = \frac{\gamma}{\varpi_c^2 R_0}\sqrt{\frac{2\pi k_B T_\alpha}{m_\alpha}}\Gamma\left(0, \frac{m_\alpha \nu_{\alpha,eff}^2 \gamma^2 R_0^2}{2k_B T_\alpha}\right)\exp\left(\frac{m_\alpha \nu_{\alpha,eff}^2 \gamma^2 R_0^2}{2k_B T_\alpha}\right), \quad (37)$$

and two radial thermal diffusion coefficients are respectively,

$$D_{T\alpha,\rho}^\rho = \frac{3\pi\gamma}{\varpi_c^2 R_0}\sqrt{\frac{2\pi k_B}{m_\alpha T_\alpha}} \; , \quad (38)$$

$$D_{T\alpha,\theta}^\rho = \sqrt{\frac{2\pi k_B}{T_\alpha m_\alpha}}\frac{\gamma}{\varpi_c^2 R_0}\left[1-\left(\frac{3}{2}+\frac{\gamma^2 R_0^2}{2k_B T_\alpha}m_\alpha \nu_{\alpha,eff}^2\right)\Gamma\left(0, \frac{m_\alpha \nu_{\alpha,eff}^2 \gamma^2 R_0^2}{2k_B T_\alpha}\right)\exp\left(\frac{m_\alpha \nu_{\alpha,eff}^2 \gamma^2 R_0^2}{2k_B T_\alpha}\right)\right] . \quad (39)$$

From Eqs.(36)-(39) we can see clearly that the four radial coefficients are all inversely proportional to square of the gyrofrequency $\varpi_c = |Q_\alpha| B / m_\alpha$, and so to square of $B$. And in Eq.(36)

and Eq.(38), dependences of the two coefficients and on the temperature $T_\alpha$ and the safe factor $\gamma$ are simple and clear. But in Eq.(37) and Eq.(39), dependences of the two coefficients and on the temperature $T_\alpha$, the effective collision frequency $v_{e,eff}$ and the safe factor $\gamma$ are very complex, and so we need to do numerical analyses.

In Eq.(30), the diffusion flux component in the $\mathbf{e}_\theta$-direction is given by

$$J_{D,\alpha}^\theta = \iiint_{\substack{0<v_{//}<\infty,\\ -\pi<\theta<\pi,\\ 0<\xi<2\pi}} \left( \frac{v_{//} \rho_g}{\gamma R_0} + v_D \cos\theta \right) f_{M,\alpha}^{(1)} dv_{//} d\theta d\xi \ . \tag{40}$$

In the same way as the calculations in Eq.(31), Eq.(40) can be calculated (see Appendix) as

$$J_{D,\alpha}^\theta = \frac{\gamma}{\varpi_c^2 R_0} \sqrt{\frac{2\pi k_B T_\alpha}{m_\alpha}} \exp\left(\frac{m_\alpha v_{\alpha,eff}^2 \gamma^2 R_0^2}{2k_B T_\alpha}\right) \Gamma\left(0, \frac{m_\alpha v_{\alpha,eff}^2 \gamma^2 R_0^2}{2k_B T_\alpha}\right) \frac{\partial n_\alpha}{\partial \rho_g}$$

$$- \frac{\gamma n_\alpha}{\varpi_c^2 R_0} \sqrt{\frac{2\pi k_B}{m_\alpha T_\alpha}} \left[ \left(\frac{3}{2} + \frac{m_\alpha v_{\alpha,eff}^2 \gamma^2 R_0^2}{2k_B T_\alpha}\right) \exp\left(\frac{m_\alpha v_{\alpha,eff}^2 \gamma^2 R_0^2}{2k_B T_\alpha}\right) \Gamma\left(0, \frac{m_\alpha v_{\alpha,eff}^2 \gamma^2 R_0^2}{2k_B T_\alpha}\right) - 1 \right] \frac{\partial T_\alpha}{\partial \rho_g}$$

$$- \left[ \frac{\rho_g^2}{v_{\alpha,eff} \gamma^2 R_0^2} + \frac{\gamma}{\varpi_c^2 R_0} \sqrt{\frac{2\pi k_B T_\alpha}{m_\alpha}} \exp\left(\frac{m_\alpha v_{\alpha,eff}^2 \gamma^2 R_0^2}{2k_B T_\alpha}\right) Erfc\sqrt{\frac{m_\alpha v_{\alpha,eff}^2 \gamma^2 R_0^2}{2k_B T_\alpha}} \right] \frac{\pi}{\rho_g} \frac{\partial n_\alpha}{\partial \theta}$$

$$- \left\{ \frac{\gamma}{\varpi_c^2 R_0} \sqrt{\frac{\pi k_B T_\alpha}{2m_\alpha}} \left[ \frac{m_\alpha v_{\alpha,eff}^2 \gamma^2 R_0^2}{k_B T_\alpha} - 3 \right] \exp\left(\frac{m_\alpha v_{\alpha,eff}^2 \gamma^2 R_0^2}{2k_B T_\alpha}\right) Erf\left(\sqrt{\frac{m_\alpha v_{\alpha,eff}^2 \gamma^2 R_0^2}{2k_B T_\alpha}}\right) \right.$$

$$\left. + \frac{v_{\alpha,eff}^2 \gamma^3 R_0}{\varpi_c^2} \left[ \frac{1}{v_{\alpha,eff} \gamma R_0} - \sqrt{\frac{\pi m_\alpha}{2k_B T_\alpha}} \exp\left(\frac{m_\alpha v_{\alpha,eff}^2 \gamma^2 R_0^2}{2k_B T_\alpha}\right) \right] \right\} \frac{\pi n_\alpha}{T_\alpha \rho_g} \frac{\partial T_\alpha}{\partial \theta} \ . \tag{41}$$

As compared with Eq.(28) we find that the angular diffusion coefficients and the angular thermal diffusion coefficients are, respectively,

$$D_{\alpha,\rho}^\theta = D_{\alpha,\theta}^\rho \ , \quad D_{T\alpha,\rho}^\theta = -D_{T\alpha,\theta}^\rho \ , \tag{42}$$

$$D_{\alpha,\theta}^\theta = \frac{\pi \rho_g^2}{v_{\alpha,eff} \gamma^2 R_0^2} + \frac{\pi \gamma}{\varpi_c^2 R_0} \sqrt{\frac{2\pi k_B T_\alpha}{m_\alpha}} \exp\left(\frac{m_\alpha v_{\alpha,eff}^2 \gamma^2 R_0^2}{2k_B T_\alpha}\right) Erfc\sqrt{\frac{m_\alpha v_{\alpha,eff}^2 \gamma^2 R_0^2}{2k_B T_\alpha}} \ , \tag{43}$$

$$D_{T\alpha,\theta}^\theta = \frac{\pi}{T_\alpha} \left\{ \frac{\gamma}{\varpi_c^2 R_0} \sqrt{\frac{\pi k_B T_\alpha}{2m_\alpha}} \left[ \frac{m_\alpha v_{\alpha,eff}^2 \gamma^2 R_0^2}{k_B T_\alpha} - 3 \right] \exp\left(\frac{m_\alpha v_{\alpha,eff}^2 \gamma^2 R_0^2}{2k_B T_\alpha}\right) Erf\left(\sqrt{\frac{m_\alpha v_{\alpha,eff}^2 \gamma^2 R_0^2}{2k_B T_\alpha}}\right) \right.$$

$$\left. + \frac{v_{\alpha,eff}^2 \gamma^3 R_0}{\varpi_c^2} \left[ \frac{1}{v_{\alpha,eff} \gamma R_0} - \sqrt{\frac{\pi m_\alpha}{2k_B T_\alpha}} \exp\left(\frac{m_\alpha v_{\alpha,eff}^2 \gamma^2 R_0^2}{2k_B T_\alpha}\right) \right] \right\} \ , \tag{44}$$

where $Erf(x)$ is the error function of $x$ and $Erfc(x)$ is the complementary error function [34].

From Eqs.(42)-(44) we can see clearly that in the four angular coefficients, only two in Eqs.(43) and (44) are independent, and they are also inversely proportional to square of the gyrofrequency $\varpi_c = |Q_\alpha| B / m_\alpha$, and so to square of $B$. But their dependences on the temperature $T_\alpha$, the effective collision frequency $v_{e,eff}$ and the safe factor $\gamma$ are complex and so need to do numerical analyses. In particular, we find that the angular diffusion coefficient in Eq.(43) is an *anomalous diffusion* coefficient, where the first term is proportional to square of the radial coordinate $\rho_g$.

In Eq.(42), $D_{\alpha,\rho}^\theta = D_{\alpha,\theta}^\rho$ means that the radial diffusion which is caused by the angular inhomogeneity is symmetric to the angular diffusion caused by the radial inhomogeneity; but $D_{T\alpha,\rho}^{\theta} = -D_{T\alpha,\theta}^{\rho}$ means that the radial thermal diffusion caused by the angular inhomogeneity is antisymmetric to the angular thermal diffusion caused by the radial inhomogeneity.

In Eq.(30), the diffusion flux component in the $\mathbf{e}_\xi$-direction is given by

$$J_{D,\alpha}^{\xi} = \iiint_{\substack{0<v_{//}<\infty,\\ -\pi<\theta<\pi,\\ 0<\xi<2\pi}} v_{//} f_{M,\alpha}^{(1)} dv_{//} d\theta d\xi \ . \tag{45}$$

In the same way as the calculations in Eq.(31), Eq.(45) can be calculated (see Appendix) as

$$J_{D,\alpha}^{\xi} = -\frac{\pi}{\gamma R_0 \nu_{\alpha,eff}} \frac{\partial n_\alpha}{\partial \theta} \ . \tag{46}$$

As compared with Eq.(28) we find the $\mathbf{e}_\xi$-direction diffusion coefficients and thermal diffusion coefficients, respectively,

$$D_{\alpha,\theta}^{\xi} = \frac{\pi \rho_g}{\gamma R_0 \nu_{\alpha,eff}} \ , \quad D_{\alpha,\rho}^{\xi} = D_{T\alpha,\rho}^{\xi} = D_{T\alpha,\theta}^{\xi} = 0 \ . \tag{47}$$

Therefore, in the $\mathbf{e}_\xi$-direction, only one coefficient is not zero, which is no relation to the temperature and the gyrofrequency, but inversely proportional to the effective collision frequency $\nu_{e,eff}$ and the safe factor $\gamma$. In particular, this diffusion coefficient in Eq.(47) is also an *anomalous diffusion* coefficient, which is proportional to the radial coordinate $\rho_g$.

Thus in the plateau regime of the toroidal plasma, we have derived the diffusion coefficient tensor and the thermal diffusion coefficient tensor, where all the matrix elements, i.e. the radial and angular diffusion coefficients as well as the radial and angular thermal diffusion coefficients, are given by Eqs.(36)-(39), Eqs.(42)-(44), and Eq.(48), respectively.

The dependences of the two radial coefficients $D_{\alpha,\rho}^{\rho}$ and $D_{T\alpha,\rho}^{\rho}$ in Eq.(36) and Eq.(38) as well as the coefficient in Eq.(47) on the temperature $T$, the gyrofrequency $\varpi_c$ and the magnetic fluid stability safety factor $\gamma$ et al are simple and clear. But, the dependences of the radial coefficients $D_{\alpha,\theta}^{\rho}$ and $D_{T\alpha,\theta}^{\rho}$ in Eqs.(37) and (39) as well as the angular coefficients in Eqs.(43)-(44) on the temperature $T_\alpha$, and the safe factor $\gamma$ and the effective collision frequency $\nu_{e,eff}$ are very complex, so we needs to do numerical analyses.

## 4 Numerical analyses of diffusion and thermal diffusion coefficients

In sec. 3, for the transport processes of charged particles in the plateau regime of the toroidal plasma, we have derived the diffusion coefficient matrix and the thermal diffusion matrix, which can be given by

$$\mathbf{D}_\alpha = \begin{pmatrix} D_{\alpha,\rho}^{\rho} & D_{\alpha,\theta}^{\rho} & 0 \\ D_{\alpha,\theta}^{\rho} & D_{\alpha,\theta}^{\theta} & 0 \\ 0 & D_{\alpha,\theta}^{\xi} & 0 \end{pmatrix}, \text{ and } \mathbf{D}_{T\alpha} = \begin{pmatrix} D_{T\alpha,\rho}^{\rho} & D_{T\alpha,\theta}^{\rho} & 0 \\ -D_{T\alpha,\theta}^{\rho} & D_{T\alpha,\theta}^{\theta} & 0 \\ 0 & 0 & 0 \end{pmatrix}, \tag{48}$$

where $D_{\alpha,\theta}^{\theta}$ and $D_{\alpha,\theta}^{\xi}$ are two *anomalous diffusion* coefficients, being a function of radial coordinate of the guiding center.

There are only seven independent matrix elements in Eq.(48). According to the previous discussions, $D_{\alpha,\rho}^{\rho}$ in Eq.(36) and $D_{T\alpha,\rho}^{\rho}$ in Eq.(38) as well as the anomalous diffusion coefficient $D_{\alpha,\theta}^{\xi}$ in Eq.(47) are simple and clear as a function of the physical quantities. The following four coefficients need to do numerical analyses so that we can show clearly their dependences on the temperature $T_\alpha$, and the magnetic fluid stability safe factor $\gamma$ and the effective collision frequency $\nu_{e,eff}$. They are the radial thermal diffusion coefficient $D_{\alpha,\theta}^{\rho}$ in Eq.(37), the radial diffusion coefficient $D_{T\alpha,\theta}^{\rho}$ in Eq.(39), the angular diffusion coefficient $D_{\alpha,\theta}^{\theta}$ in Eq.(43), and the angular thermal diffusion coefficient $D^{\theta}_{T\alpha,\theta}$ in Eq.(44). In particular, we need to analyze the properties of the anomalous angular diffusion coefficient $D^{\theta}_{\alpha,\theta}$.

Here we take an electron in the plasma as an example to do the numerical analyses. From Ref.[2], the basic physical data of the toroidal plasma for the numerical analysis are listed in Table 1.

Table 1: The basic physical data in the toroidal plasma [2] for the numerical analysis

| Electronic number density | $n_e$ | ~$10^{20}$m$^{-3}$ |
|---|---|---|
| Aspect ratio | $\varepsilon$ | 0.2 |
| Electron collision frequency | $v_{ee}$ | ~$5.27 \times 10^3$Hz |
| Effective collision frequency | $v_{e,eff}$ (=$v_{ee}/\varepsilon$) | ~$2.64 \times 10^4$Hz |
| Radius of the toroidal magnetic field | $R_0$ | 3m |
| Magnetic fluid stability safety factor | $\gamma$ | 1.5 |
| Magnetic induction intensity | $B$ | 3.5 T |
| Edge radius of magnetic field | $a$ | $a=\varepsilon R_0$=0.6m |
| Plasma temperature | $T_e$ | ~10keV=$1.16 \times 10^8$K |

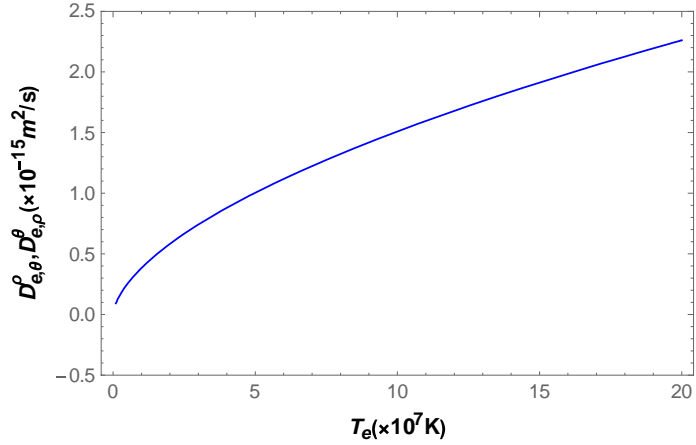

Fig. 1(a). Dependence of the diffusion coefficient $D^{\rho}_{e,\theta}$ ( $D^{\theta}_{e,\rho}$ ) on temperature $T_e$.

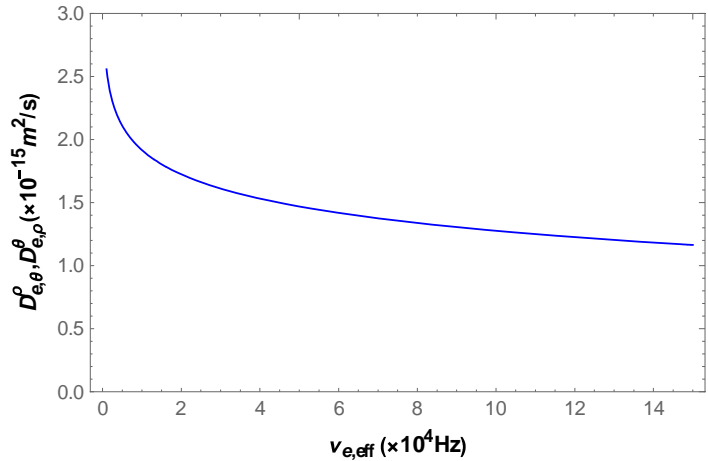

Fig. 1(b). Dependence of the diffusion coefficient $D^{\rho}_{e,\theta}$ ( $D^{\theta}_{e,\rho}$ ) on the effective collision frequency $v_{e,eff}$.

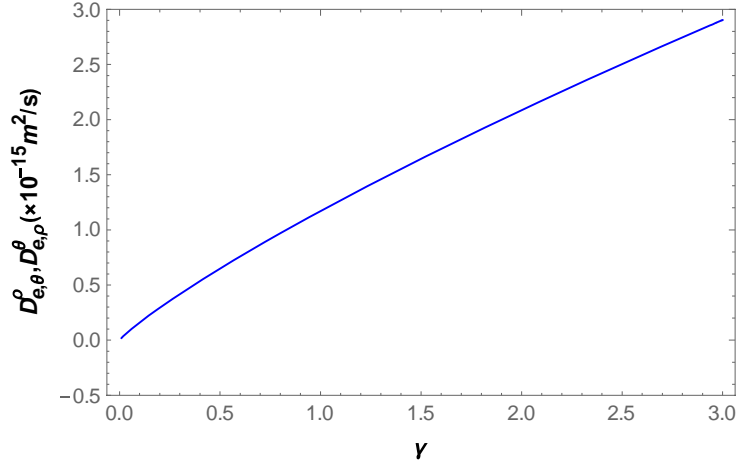

Fig. 1(c). Dependence of the diffusion coefficient $D^{\rho}_{e,\theta}$ ( $D^{\theta}_{e,\rho}$ ) on the magnetic fluid stability factor $\gamma$ .

In Figs. 1(a)-1(c), we give the numerical results of the dependences of the radial diffusion coefficient $D^{\rho}_{e,\theta}$ (or the angular diffusion coefficient $D^{\theta}_{e,\rho}$) on temperature $T_e$, on the effective collision frequency $v_{e,eff}$ and on the magnetic fluid stability factor $\gamma$, respectively.

In Fig.1(a), we show that this diffusion coefficient increases monotonously with the increase of temperature $T_e$, but in Fig.1(b) it decreases rapidly with the increase of effective collision frequency at the lower frequency (e.g., $v_{e,eff} < 2 \times 10^4 Hz$ ) and as the effective collision frequency continues to increase, it is decreasing slowly to approach a constant approximately. In Fig.1(c), this diffusion coefficient increases almost linearly with the increase of $\gamma$, which means that this radial diffusion caused by the angular inhomogeneity (or this angular diffusion caused by the radial inhomogeneity) can increase almost proportional to the magnetic fluid stability factor $\gamma$.

In Figs. 2 (a)-2(c), for four different radial positions, $\rho_g$= 0, $\rho_g$=0.3$a$, $\rho_g$=0.6$a$ and $\rho_g$=$a$, we give the numerical results of the dependences of the anomalous angular diffusion coefficient $D^{\theta}_{e,\theta}$ on the temperature $T_e$ , on the effective collision frequency $v_{e,eff}$ and on the magnetic fluid stability factor $\gamma$, respectively.

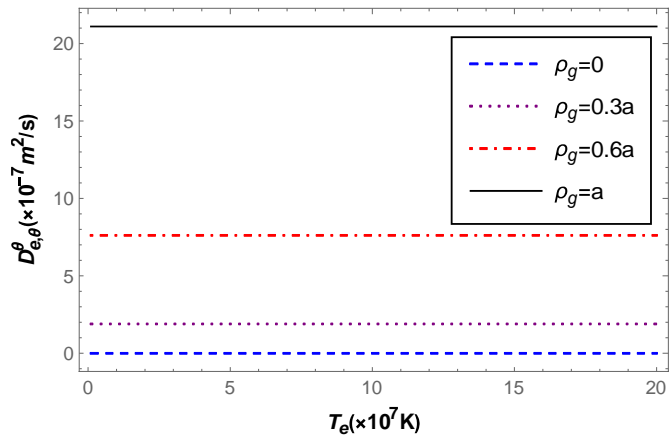

Fig. 2(a). Dependence of the anomalous diffusion coefficient $D^{\theta}_{e,\theta}$ on temperature $T_e$.

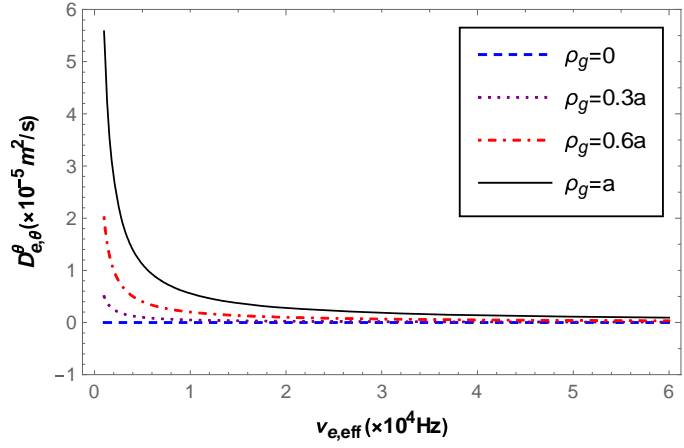

Fig. 2(b). Dependence of the anomalous diffusion coefficient $D^{\theta}_{e,\theta}$ on the effective collision frequency $v_{e,eff}$.

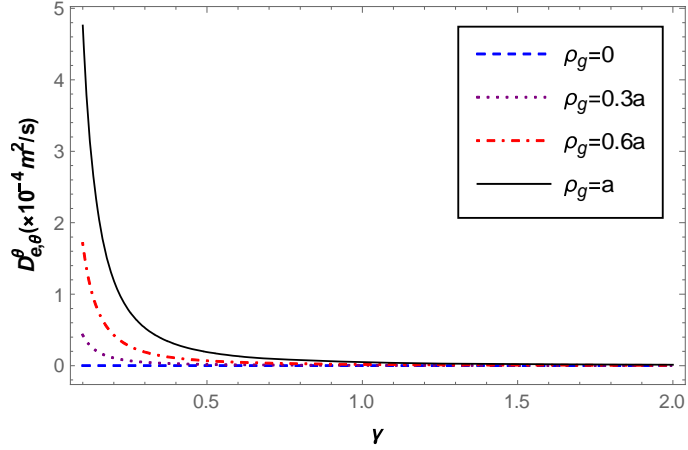

Fig. 2(c). Dependence of the anomalous diffusion coefficient $D^{\theta}_{e,\theta}$ on the magnetic fluid stability factor $\gamma$.

In Fig. 2(a), we show that the anomalous angular diffusion coefficient $D^{\theta}_{e,\theta}$ is basically a constant with the change of temperature $T_e$ for the different $\rho_g$. Therefore, it can be considered to be independent of temperature within the temperature range of fusion plasma. In Figs. 2(b) and 2(c), we show that this anomalous diffusion coefficient decreases rapidly with the increase of the effective collision frequency $v_{e,eff}$ or with the increase of the magnetic fluid stability factor $\gamma$, and it approaches to zero basically when the effective collision frequency $v_{e,eff}$ is $v_{e,eff} > 5 \times 10^4 Hz$ or the magnetic fluid stability factor is $\gamma>1$ approximately, or the radial position is $\rho_g=0$. Therefore, the anomalous diffusion coefficient $D^{\theta}_{e,\theta}$ depends significantly on the radial position $\rho_g$, but has nothing to do with the second term in Eq.(45) because it is zero basically. Based on the numerical analyses, the anomalous angular diffusion coefficient in Eq.(45) can be reduced for the present toroidal plasma to be

$$D^{\theta}_{\alpha,\theta} \approx \frac{\pi \rho_g^2}{v_{\alpha,eff} \gamma^2 R_0^2} \ . \tag{49}$$

In Figs. 3(a)-3(c), we give the numerical results of the dependences of the angular thermal diffusion coefficient $D^{\theta}_{Te,\rho}$ (or the radial thermal diffusion coefficient $-D^{\rho}_{Te,\theta}$) on the temperature $T_e$, the effective collision frequency $v_{e,eff}$ and the magnetic fluid stability factor $\gamma$, respectively.

In Figs. 3(a) we show that the angular thermal diffusion coefficient $D^{\theta}_{Te,\rho}$ decrease rapidly with increase of the temperature $T_e$ when the temperature $T_e$ is lower (e.g., $T_e < 2 \times 10^7 K$), then it decreases slowly with increase of the temperature $T_e$ when the temperature becomes higher (e.g., $T_e > 2 \times 10^7 K$ ), and approaches to a constant gradually. In Fig. 3(b) we show that the thermal diffusion coefficient $D_{Te,\rho}^{\theta}$ decreases monotonically with the increase of the effective collision frequency $v_{e,eff}$, but the decreasing rate slows down gradually as $v_{e,eff}$ increases. In Fig. 3(c) we show effect of the magnetic fluid stability factor $\gamma$ on the thermal diffusion coefficient $D_{Te,\rho}^{\theta}$. It is clear that this thermal diffusion coefficient increase almost linearly with increase of the magnetic fluid stability factor $\gamma$. It is worth noting that because of $D_{Te,\rho}^{\theta} = -D_{Te,\theta}^{\rho}$, the radial thermal diffusion caused by the angular inhomogeneity has the same properties as the angular thermal diffusion caused by the radial inhomogeneity.

In Figs. 4(a)-4(c), we give the numerical results of the dependences of the angular thermal diffusion coefficient $D_{Te,\theta}^{\theta}$ on the temperature $T_e$, the effective collision frequency $v_{e,eff}$ and the magnetic fluid stability factor $\gamma$, respectively. We show that this thermal diffusion coefficient is negative.

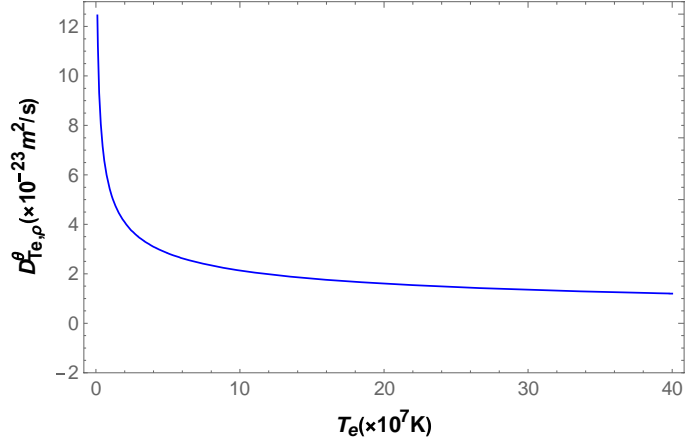

Fig. 3(a). Dependence of the thermal diffusion coefficient $D_{Te,\rho}^{\theta}$ ( $-D_{Te,\theta}^{\rho}$ ) on the temperature $T_e$.

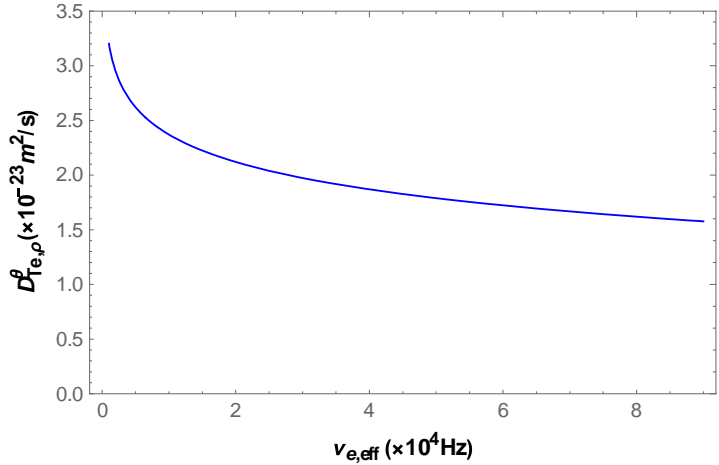

Fig. 3(b). Dependence of the thermal diffusion coefficient $D_{Te,\rho}^{\theta}$ ( $-D_{Te,\theta}^{\rho}$ ) on the effective collision frequency $v_{e,eff}$.

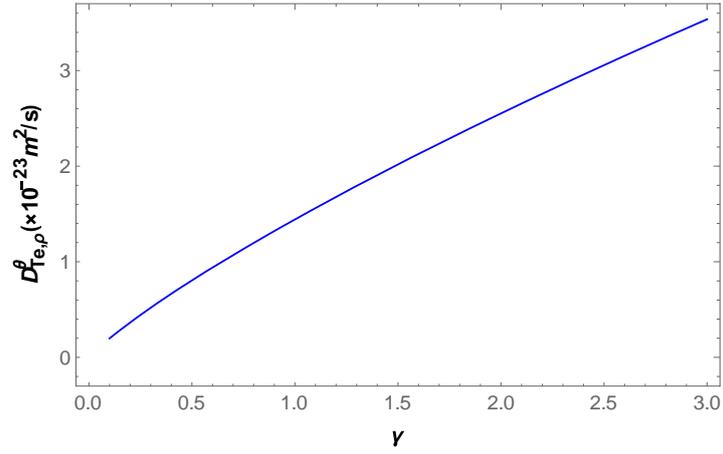

Fig. 3(c). Dependence of the thermal diffusion coefficient $D^{\theta}_{Te,\rho}$ ( $-D^{\rho}_{Te,\theta}$ ) on the magnetic fluid stability factor $\gamma$.

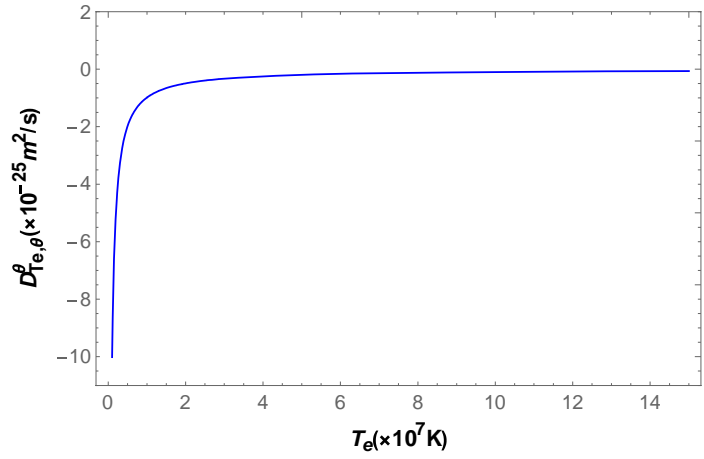

Fig. 4(a). Dependence of the thermal diffusion coefficient $D^{\theta}_{Te,\theta}$ on the temperature $T_e$.

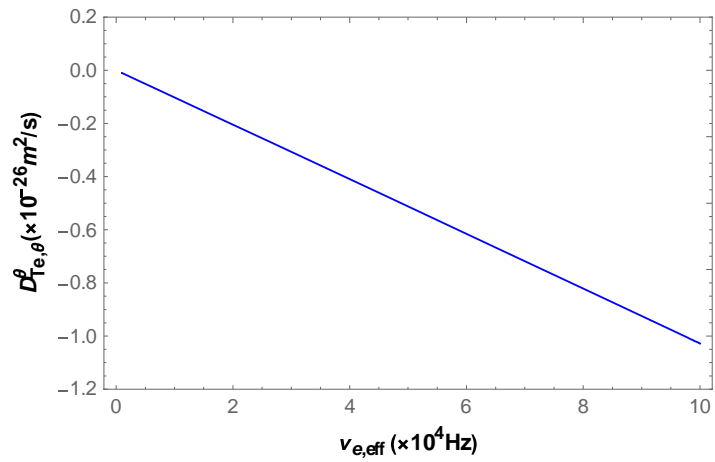

Fig. 4(b). Dependence of the thermal diffusion coefficient $D^{\theta}_{Te,\theta}$ on the effective collision frequency $v_{e,eff}$.

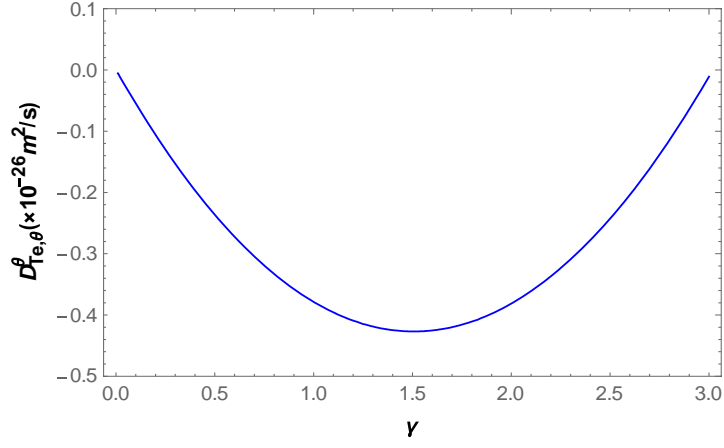

Fig. 4(c). Dependence of the thermal diffusion coefficient $D^{\theta}_{Te,\theta}$ on the magnetic fluid stability factor $\gamma$.

In Fig. 4(a) we show that the angular thermal diffusion coefficient $D^{\theta}_{Te,\theta}$ increases very rapidly with increase of the temperature (i.e., its absolute value increases) when $T_e$ is lower (viz., $T_e < 10^7 K$), but as the temperature continues to increase (e.g., $T_e > 10^7 K$), the increase of $D^{\theta}_{Te,\theta}$ becomes slow and it will approach to a constant close to zero. In Fig. 4(b), we show that this thermal diffusion coefficient decreases almost linearly with increase of the effective collision frequency $v_{e,eff}$ (i.e., its absolute value increases almost linearly). In Fig. 4(c), we show that this thermal diffusion coefficient decreases (i.e., its absolute value increases) with increase of the magnetic fluid stability factor $\gamma$ when $\gamma$ is smaller, and then it approaches the minimum value (its absolute value approaches the maximum value) at $\gamma \approx 1.5$ approximately, but it will increase (its absolute value will decrease) with increase of $\gamma$ when $\gamma$ is larger (e.g., $\gamma > 1.5$).

## 5 Conclusions

In conclusion, we have studied the diffusions and anomalous diffusions of charged particles in the plateau regime of the toroidal plasma confined by a toroidal magnetic field. As usual, the fundamental equilibrium velocity distribution function is based on a Maxwell-Boltzmann velocity distribution. By using the kinetic transport equation, we derived the first-order approximation of Chapman-Enskog expansion for the velocity distribution function of charged particles in the plateau regime of toroidal plasma, and it was given under the toroidal coordinate reference system.

Based on the velocity distribution function of charged particles in the plateau regime of toroidal plasma under the toroidal coordinate reference system, we have derived the diffusion flux vector, and then derived the diffusion coefficient tensor and the thermal diffusion coefficient tensor. Therefore, we obtained the diffusion coefficients and the thermal diffusion coefficients of charged particles in the plateau regime of the toroidal plasma under a 3-dimensional toroidal coordinate system, given by Eqs. (36)-(39), Eqs.(42)-(44) and Eq.(47), respectively, where the angular diffusion coefficient $D^{\theta}_{\alpha,\theta}$ in Eq.(43) and $D^{\xi}_{\alpha,\theta}$ in Eq.(47) are two *anomalous diffusions* because they depend on the radial position coordinates $\rho_g$.

In the diffusion matrix elements, only four diffusion coefficients are independent, where there is $D^{\rho}_{\alpha,\theta} = D^{\theta}_{\alpha,\rho}$, i.e., the radial diffusion that is caused by the angular inhomogeneity is symmetric to the angular diffusion that is caused by the radial inhomogeneity. But in the thermal diffusion matrix elements, only three thermal diffusion coefficients are independent, where there is $D^\rho_{T\alpha,\theta} = -D^\theta_{T\alpha,\rho}$, i.e., the radial thermal diffusion that is caused by the angular inhomogeneity is anti-symmetric to the angular thermal diffusion that is caused by the radial inhomogeneity.

In the seven independent coefficients, the radial diffusion coefficient $D^\rho_{\alpha,\rho}$ in Eq.(36), the radial thermal diffusion coefficient $D^\rho_{T\alpha,\rho}$ in Eq.(38) and the $\mathbf{e}_\xi$-direction anomalous diffusion coefficient $D^\xi_{\alpha,\theta}$ in Eq.(47) are all simple and clear as a function of the temperature $T_\alpha$, the gyrofrequency $\varpi_c$, the effective collision frequency $\nu_{e,eff}$, the magnetic fluid stability factor $\gamma$ and the radial position coordinates $\rho_g$, respectively, i.e.,

$$D^\rho_{\alpha,\rho} = \frac{2\pi\gamma}{\varpi_c^2 R_0}\sqrt{\frac{2\pi k_B T_\alpha}{m_\alpha}}, \quad D^\rho_{T\alpha,\rho} = \frac{3\pi\gamma}{\varpi_c^2 R_0}\sqrt{\frac{2\pi k_B}{m_\alpha T_\alpha}}, \quad D^\xi_{\alpha,\theta} = \frac{\pi\rho_g}{\gamma R_0 \nu_{\alpha,eff}}. \tag{50}$$

But the other four independent coefficients, the radial thermal diffusion coefficient $D^\rho_{\alpha,\theta}$ in Eq.(37), the radial diffusion coefficient $D^\rho_{T\alpha,\theta}$ in Eq.(39), the angular *anomalous diffusion* coefficient $D^\theta_{\alpha,\theta}$ in Eq.(43), and the angular thermal diffusion coefficient $D^\theta_{T\alpha,\theta}$ in Eq.(44), are complex as a function of the above physical quantities, so we have made the numerical analyses of them by taking electrons as an example.

The numerical results have been given in Figs. 1(a)-(c), Figs. 2(a)-(c), Figs. 3(a)-(c), and Figs. 4(a)-(c), respectively, which have shown clearly their properties as a function of the temperature $T_\alpha$, the effective collision frequency $\nu_{e,eff}$, the magnetic fluid stability factor $\gamma$ and the radial position coordinates $\rho_g$, respectively. And by the numerical analyses we the second term in Eq.(43) can be neglected for the present toroidal plasma and thus the angular *anomalous diffusion* coefficient $D^\theta_{\alpha,\theta}$ can be expressed by Eq.(49) approximately.

It is also shown that, except the $\mathbf{e}_\xi$-direction anomalous diffusion coefficient $D^\xi_{\alpha,\theta}$ ( it has no relation with the gyrofrequency), the six independent coefficients are all inversely proportional to square of the gyrofrequency $\varpi_c = |Q_\alpha| B / m_\alpha$, and so to square of the magnetic induction $B$.

**Acknowledgements** This work is supported by the National Natural Science Foundation of China under Grant No. 11775156.

**Appendix**

After doing the approximation and simplification in Eqs.(32)-(34), Eq.(31) can be calculated by

$$\begin{aligned}
J^\rho_{\alpha,D} &= -\iiint_{\substack{0<v_\parallel<\infty,\\-\pi<\theta<\pi,\\0<\xi<2\pi}} v_D^2 \sin^2\theta \frac{1}{\rho_g}\frac{\partial f^{(0)}_{M,\alpha}}{\partial\theta}\frac{\frac{v_\parallel}{\gamma R_0}}{\nu_{\alpha,eff}^2 + \frac{v_\parallel^2}{\gamma^2 R_0^2}}dv_\parallel d\theta d\xi - \iiint_{\substack{0<v_\parallel<\infty,\\-\pi<\theta<\pi,\\0<\xi<2\pi}} v_D^2 \sin^2\theta \frac{\partial f^{(0)}_{M,\alpha}}{\partial\rho_g}\frac{\nu_{\alpha,eff}}{\nu_{\alpha,eff}^2 + \frac{v_\parallel^2}{\gamma^2 R_0^2}}dv_\parallel d\theta d\xi \\
&= -2\pi^2\left(\frac{2k_B T_\alpha}{m_\alpha \varpi_c R_0}\right)^2 \frac{1}{\rho_g}\int_0^\infty \frac{\partial f^{(0)}_{M,\alpha}}{\partial\theta}\frac{\frac{v_\parallel}{\gamma R_0}}{\nu_{\alpha,eff}^2 + \frac{v_\parallel^2}{\gamma^2 R_0^2}}dv_\parallel - 2\pi^2\left(\frac{2k_B T_\alpha}{m_\alpha \varpi_c R_0}\right)^2 \int_0^\infty \frac{\partial f^{(0)}_{M,\alpha}}{\partial\rho_g}\pi\delta\left(\frac{v_\parallel}{\gamma R_0}\right)dv_\parallel \\
&= -2\left(\frac{2\pi k_B T_\alpha}{m_\alpha \varpi_c R_0}\right)^2 \Bigg[\frac{1}{\rho_g}\int_0^\infty \left(\frac{1}{n_\alpha}\frac{\partial n_\alpha}{\partial\theta} - \frac{3}{2T_\alpha}\frac{\partial T_\alpha}{\partial\theta} + \frac{m_\alpha v_\parallel^2}{2k_B T_\alpha^2}\frac{\partial T_\alpha}{\partial\theta}\right)\frac{\frac{v_\parallel}{\gamma R_0}f^{(0)}_{M,\alpha}}{\nu_{\alpha,eff}^2 + \frac{v_\parallel^2}{\gamma^2 R_0^2}}dv_\parallel \\
&\quad + \gamma R_0 \int_0^\infty \left(\frac{1}{n_\alpha}\frac{\partial n_\alpha}{\partial\rho_g} - \frac{3}{2T_\alpha}\frac{\partial T_\alpha}{\partial\rho_g} + \frac{m_\alpha v_\parallel^2}{2k_B T_\alpha^2}\frac{\partial T_\alpha}{\partial\rho_g}\right)f^{(0)}_{M,\alpha}\delta(v_\parallel)dv_\parallel\Bigg]
\end{aligned}$$

$$= -\left(\frac{1}{\varpi_c R_0}\right)^2 \frac{2}{\rho_g} \sqrt{\frac{2\pi k_B T_\alpha}{m_\alpha}} \int_0^\infty e^{-\frac{m_\alpha v_{//}^2}{2k_B T_\alpha}} \left(\frac{\partial n_\alpha}{\partial \theta} - \frac{3n_\alpha}{2T_\alpha}\frac{\partial T_\alpha}{\partial \theta} + n_\alpha \frac{m_\alpha v_{//}^2}{2k_B T_\alpha^2}\frac{\partial T_\alpha}{\partial \theta}\right) \frac{\frac{v_{//}}{\gamma R_0}}{v_{\alpha,eff}^2 + \frac{v_{//}^2}{\gamma^2 R_0^2}} dv_{//}$$

$$-2\pi\gamma R_0 \left(\frac{1}{\varpi_c R_0}\right)^2 \sqrt{\frac{2\pi k_B T_\alpha}{m_\alpha}} \int_0^\infty e^{-\frac{m_\alpha v_{//}^2}{2k_B T_\alpha}} \left(\frac{\partial n_\alpha}{\partial \rho_g} - \frac{3n_\alpha}{2T_\alpha}\frac{\partial T_\alpha}{\partial \rho_g}\right) \delta(v_{//}) dv_{//}$$

$$= -2\pi^{3/2} \sqrt{\frac{2k_B T_\alpha}{m_\alpha}} \frac{\gamma}{\varpi_c^2 R_0} \frac{\partial n_\alpha}{\partial \rho_g} + 3\pi \sqrt{\frac{2\pi k_B T_\alpha}{m_\alpha}} \frac{\gamma}{\varpi_c^2 R_0} \frac{n_\alpha}{T_\alpha} \frac{\partial T_\alpha}{\partial \rho_g}$$

$$-\sqrt{\frac{2\pi k_B T_\alpha}{m_\alpha}} \frac{\gamma}{\varpi_c^2 R_0} \Gamma\left(0, \frac{m_\alpha v_{\alpha,eff}^2 \gamma^2 R_0^2}{2k_B T_\alpha}\right) \exp\left(\frac{m_\alpha v_{\alpha,eff}^2 \gamma^2 R_0^2}{2k_B T_\alpha}\right) \frac{1}{\rho_g} \frac{\partial n_\alpha}{\partial \theta}$$

$$-\sqrt{\frac{2\pi k_B}{m_\alpha T_\alpha}} \frac{\gamma}{2\varpi_c^2 R_0} \left[2 - \left(3 + \frac{\gamma^2 R_0^2}{k_B T_\alpha} m_\alpha v_{\alpha,eff}^2\right) \Gamma\left(0, \frac{m_\alpha v_{\alpha,eff}^2 \gamma^2 R_0^2}{2k_B T_\alpha}\right) \exp\left(\frac{m_\alpha v_{\alpha,eff}^2 \gamma^2 R_0^2}{2k_B T_\alpha}\right)\right] \frac{n_\alpha}{\rho_g} \frac{\partial T_\alpha}{\partial \theta}. \quad (A.1)$$

This is Eq.(35).

In the same way as the calculations in Eq.(31), Eq.(39) can be calculated by

$$\mathbf{J}_{\alpha,D}^\theta = -\iiint_{\substack{0<v_{//}<\infty,\\ -\pi<\theta<\pi,\\ 0<\xi<2\pi}} \frac{v_{//}}{\gamma R_0} \frac{1}{v_{\alpha,eff}} \left(\frac{v_{//}\rho_g}{\gamma R_0} + v_D \cos\theta\right) \frac{\partial f_{M,\alpha}^{(0)}}{\partial \theta} dv_{//} d\theta d\xi$$

$$-\iiint_{\substack{0<v_{//}<\infty,\\ -\pi<\theta<\pi,\\ 0<\xi<2\pi}} \left(v_D^2 \cos^2\theta + \frac{v_{//}\rho_g}{\gamma R_0} v_D \cos\theta\right) \frac{-\frac{\partial f_{M,\alpha}^{(0)}}{\partial \rho_g}\frac{v_{//}}{\gamma R_0} + \frac{\partial f_{M,\alpha}^{(0)}}{\partial \theta}\frac{v_{\alpha,eff}}{\rho_g}}{v_{\alpha,eff}^2 + \frac{v_{//}^2}{\gamma^2 R_0^2}} dv_{//} d\theta d\xi$$

$$-\iiint_{\substack{0<v_{//}<\infty,\\ -\pi<\theta<\pi,\\ 0<\xi<2\pi}} \left(\frac{v_{//}\rho_g}{\gamma R_0} v_D \sin\theta + v_D^2 \sin\theta\cos\theta\right) \frac{\frac{\partial f_{M,\alpha}^{(0)}}{\partial \theta}\frac{v_{//}}{\gamma R_0}\frac{1}{\rho_g} + \frac{\partial f_{M,\alpha}^{(0)}}{\partial \rho_g}v_{\alpha,eff}}{v_{\alpha,eff}^2 + \frac{v_{//}^2}{\gamma^2 R_0^2}} dv_{//} d\theta d\xi$$

$$= -\iiint_{\substack{0<v_{//}<\infty,\\ -\pi<\theta<\pi,\\ 0<\xi<2\pi}} \frac{\rho_g v_{//}^2}{v_{\alpha,eff} \gamma^2 R_0^2} \frac{\partial f_{M,\alpha}^{(0)}}{\partial \theta} dv_{//} d\theta d\xi - \iiint_{\substack{0<v_{//}<\infty,\\ -\pi<\theta<\pi,\\ 0<\xi<2\pi}} v_D^2 \cos^2\theta \frac{-\frac{\partial f_{M,\alpha}^{(0)}}{\partial \rho_g}\frac{v_{//}}{\gamma R_0} + \frac{\partial f_{M,\alpha}^{(0)}}{\partial \theta}\frac{v_{\alpha,eff}}{\rho_g}}{v_{\alpha,eff}^2 + \frac{v_{//}^2}{\gamma^2 R_0^2}} dv_{//} d\theta d\xi$$

$$= -\frac{4\pi^2 \rho_g}{\gamma^2 R_0^2 v_{\alpha,eff}} \int_0^\infty v_{//}^2 \frac{\partial f_{M,\alpha}^{(0)}}{\partial \theta} dv_{//} + 2\pi^2 \int_0^\infty v_D^2 \frac{\partial f_{M,\alpha}^{(0)}}{\partial \rho_g} \frac{\frac{v_{//}}{\gamma R_0}}{v_{\alpha,eff}^2 + \frac{v_{//}^2}{\gamma^2 R_0^2}} dv_{//} - 2\pi^2 \frac{v_{\alpha,eff}}{\rho_g} \int_0^\infty v_D^2 \frac{\partial f_{M,\alpha}^{(0)}}{\partial \theta} \frac{dv_{//}}{v_{\alpha,eff}^2 + \frac{v_{//}^2}{\gamma^2 R_0^2}}$$

$$= -\frac{4\pi^2 \rho_g}{v_{\alpha,eff} \gamma^2 R_0^2} \left(\frac{m_\alpha}{2\pi k_B T_\alpha}\right)^{\frac{3}{2}} \int_0^\infty \left(\frac{\partial n_\alpha}{\partial \theta} - \frac{3n_\alpha}{2T_\alpha}\frac{\partial T_\alpha}{\partial \theta} + n_\alpha \frac{m_\alpha v_{//}^2}{2k_B T_\alpha^2}\frac{\partial T_\alpha}{\partial \theta}\right) v_{//}^2 e^{-\frac{m_\alpha v_{//}^2}{2k_B T_\alpha}} dv_{//}$$

$$+ \left(\frac{1}{\varpi_c R_0}\right)^2 \sqrt{\frac{8\pi k_B T_\alpha}{m_\alpha}} \int_0^\infty \left(\frac{\partial n_\alpha}{\partial \rho_g} - \frac{3n_\alpha}{2T_\alpha}\frac{\partial T_\alpha}{\partial \rho_g} + n_\alpha \frac{m_\alpha v_{//}^2}{2k_B T_\alpha^2}\frac{\partial T_\alpha}{\partial \rho_g}\right) e^{-\frac{m_\alpha v_{//}^2}{2k_B T_\alpha}} \frac{\frac{v_{//}}{\gamma R_0}}{v_{\alpha,eff}^2 + \frac{v_{//}^2}{\gamma^2 R_0^2}} dv_{//}$$

$$-\frac{v_{\alpha,eff}}{\rho_g} \left(\frac{1}{\varpi_c R_0}\right)^2 \sqrt{\frac{8\pi k_B T_\alpha}{m_\alpha}} \int_0^\infty \left(\frac{\partial n_\alpha}{\partial \theta} - \frac{3n_\alpha}{2T_\alpha}\frac{\partial T_\alpha}{\partial \theta} + n_\alpha \frac{m_\alpha v_{//}^2}{2k_B T_\alpha^2}\frac{\partial T_\alpha}{\partial \theta}\right) e^{-\frac{m_\alpha v_{//}^2}{2k_B T_\alpha}} \frac{dv_{//}}{v_{\alpha,eff}^2 + \frac{v_{//}^2}{\gamma^2 R_0^2}}$$

$$= \frac{\gamma}{\varpi_c^2 R_0} \sqrt{\frac{2\pi k_B T_\alpha}{m_\alpha}} \exp\left(\frac{m_\alpha v_{\alpha,eff}^2 \gamma^2 R_0^2}{2 k_B T_\alpha}\right) \Gamma\left(0, \frac{m_\alpha v_{\alpha,eff}^2 \gamma^2 R_0^2}{2 k_B T_\alpha}\right) \frac{\partial n_\alpha}{\partial \rho_g}$$

$$- \frac{\gamma n_\alpha}{\varpi_c^2 R_0} \sqrt{\frac{2\pi k_B}{m_\alpha T_\alpha}} \left[\left(\frac{3}{2} + \frac{m_\alpha v_{\alpha,eff}^2 \gamma^2 R_0^2}{2 k_B T_\alpha}\right) \exp\left(\frac{m_\alpha v_{\alpha,eff}^2 \gamma^2 R_0^2}{2 k_B T_\alpha}\right) \Gamma\left(0, \frac{m_\alpha v_{\alpha,eff}^2 \gamma^2 R_0^2}{2 k_B T_\alpha}\right) - 1\right] \frac{\partial T_\alpha}{\partial \rho_g}$$

$$- \left[\frac{\rho_g^2}{v_{\alpha,eff} \gamma^2 R_0^2} + \frac{\gamma}{\varpi_c^2 R_0} \sqrt{\frac{2\pi k_B T_\alpha}{m_\alpha}} \exp\left(\frac{m_\alpha v_{\alpha,eff}^2 \gamma^2 R_0^2}{2 k_B T_\alpha}\right) Erfc \sqrt{\frac{m_\alpha v_{\alpha,eff}^2 \gamma^2 R_0^2}{2 k_B T_\alpha}}\right] \frac{\pi}{\rho_g} \frac{\partial n_\alpha}{\partial \theta}$$

$$- \left\{\frac{\gamma}{\varpi_c^2 R_0} \sqrt{\frac{\pi k_B T_\alpha}{2 m_\alpha}} \left[\frac{m_\alpha v_{\alpha,eff}^2 \gamma^2 R_0^2}{k_B T_\alpha} - 3\right] \exp\left(\frac{m_\alpha v_{\alpha,eff}^2 \gamma^2 R_0^2}{2 k_B T_\alpha}\right) Erf\left(\sqrt{\frac{m_\alpha v_{\alpha,eff}^2 \gamma^2 R_0^2}{2 k_B T_\alpha}}\right) \right.$$

$$\left. + \frac{v_{\alpha,eff}^2 \gamma^3 R_0}{\varpi_c^2} \left[\frac{1}{v_{\alpha,eff} \gamma R_0} - \sqrt{\frac{\pi m_\alpha}{2 k_B T_\alpha}} \exp\left(\frac{m_\alpha v_{\alpha,eff}^2 \gamma^2 R_0^2}{2 k_B T_\alpha}\right)\right]\right\} \frac{\pi n_\alpha}{T_\alpha \rho_g} \frac{\partial T_\alpha}{\partial \theta}. \quad \text{(A.2)}$$

This is Eq.(40).

In the same way as the calculations in Eq.(31), Eq.(45) can be calculated by

$$J_{\alpha,D}^\xi = -\iiint_{\substack{0<v_{//}<\infty,\\-\pi<\theta<\pi,\\0<\xi<2\pi}} \frac{v_{//}^2}{\gamma R_0} \frac{1}{v_{\alpha,eff}} \frac{\partial f_{M,\alpha}^{(0)}}{\partial \theta} dv_{//} d\theta d\xi - \iiint_{\substack{0<v_{//}<\infty,\\-\pi<\theta<\pi,\\0<\xi<2\pi}} v_{//} v_D \cos\theta \frac{-\frac{v_{//}}{\gamma R_0} \frac{\partial f_{M,\alpha}^{(0)}}{\partial \rho_g} + \frac{v_{\alpha,eff}}{\rho_g} \frac{\partial f_{M,\alpha}^{(0)}}{\partial \theta}}{v_{\alpha,eff}^2 + \frac{v_{//}^2}{\gamma^2 R_0^2}} dv_{//} d\theta d\xi$$

$$- \iiint_{\substack{0<v_{//}<\infty,\\-\pi<\theta<\pi,\\0<\xi<2\pi}} v_{//} v_D \sin\theta \frac{\frac{v_{//}}{\gamma R_0} \frac{1}{\rho_g} \frac{\partial f_{M,\alpha}^{(0)}}{\partial \theta} + v_{\alpha,eff} \frac{\partial f_{M,\alpha}^{(0)}}{\partial \rho_g}}{v_{\alpha,eff}^2 + \frac{v_{//}^2}{\gamma^2 R_0^2}} dv_{//} d\theta d\xi$$

$$= -\frac{1}{\gamma R_0 v_{\alpha,eff}} \iiint_{\substack{0<v_{//}<\infty,\\-\pi<\theta<\pi,\\0<\xi<2\pi}} v_{//}^2 \frac{\partial f_{M,\alpha}^{(0)}}{\partial \theta} dv_{//} d\theta d\xi$$

$$= -\frac{4\pi^2}{\gamma R_0 v_{\alpha,eff}} \left(\frac{m_\alpha}{2\pi k_B T_\alpha}\right)^{\frac{3}{2}} \int_0^\infty \left(\frac{\partial n_\alpha}{\partial \theta} - \frac{3 n_\alpha}{2 T_\alpha} \frac{\partial T_\alpha}{\partial \theta} + n_\alpha \frac{m_\alpha v_{//}^2}{2 k_B T_\alpha^2} \frac{\partial T_\alpha}{\partial \theta}\right) v_{//}^2 e^{-\frac{m_\alpha v_{//}^2}{2 k_B T_\alpha}} dv_{//}$$

$$= -\frac{\pi}{\gamma R_0 v_{\alpha,eff}} \frac{\partial n_\alpha}{\partial \theta}. \quad \text{(A.3)}$$

This is Eq.(46).